\documentclass[12pt]{amsart}
\usepackage{latexsym,amsmath,amsfonts,amscd,amssymb}
\usepackage{graphicx}
\newtheorem{theorem*}{Theorem}

\newtheorem{proposition*}[theorem*]{Proposition}

\newtheorem{corollary*}[theorem*]{Corollary}

\theoremstyle{remark}

\newtheorem{remark*}[theorem*]{Remark}

\newtheorem{note*}[theorem*]{Note}

\title{Ant routing algorithm for the Lightning Network}

\subjclass[2010]{}

\author[C. Grunspan]{Cyril Grunspan}
\address{Cyril Grunspan\newline{}\indent L\'eonard de Vinci P\^ole Univ, Research Center, Labex R\'efi, \newline{}\indent 
Paris, France, }
\email{cyril.grunspan@devinci.fr}
\author[R. P\'{e}rez-Marco]{Ricardo P\'{e}rez-Marco}
\address{Ricardo P\'{e}rez-Marco\newline{}\indent CNRS, IMJ-PRG, Univ. Paris 7, Labex R\'efi, \newline{}\indent 
Paris, France}
\email{ricardo.perez.marco@gmail.com}
\address{\tiny Author's Bitcoin Beer Address (ABBA)\footnote{\tiny Send some bitcoins to support our research at the pub.}:
1KrqVxqQFyUY9WuWcR5EHGVvhCS841LPLn} 

\address{\includegraphics[scale=0.6]{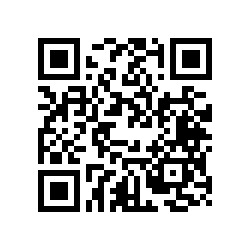}}

\begin{document}

\begin{abstract}
  We propose a decentralized routing algorithm that can be implemented in Bitcoin Lightning Network. 
  All nodes in the network contribute equally to path searching. The algorithm is inspired from  
  ant path searching algorithms.
\end{abstract}

{\maketitle}

\section{Introduction}

Bitcoin's network is a decentralized peer-to-peer payment network \cite{N08} that allows programmable 
transactions. Transactions are not instantaneous, and the 
current protocol does not scale to thousands of transactions per second as it will be necessary 
to scale to a worldwide payment system. The Lightning Network is a second layer payment network with 
these properties. Security of the Lightning Network relies on the Bitcoin network. We want to build such 
an extension preserving decentralization and confidentiality of transactions.

\medskip

Bitcoin scripting language (and proposed extensions) allows to open one way or bidirectional 
payment channels (\cite{DW15}, \cite{HS15}, \cite{PD15}, \cite{DOR18})  so that 
two individuals can perform peer-to-peer
almost instantaneous, anonymous and secure off-chain transactions. 
Only the Initial Commitment Transaction and the final Settlement Transaction need to be 
registered in Bitcoin's blockchain. 

\medskip

Hashed Timelocked Contracts (HTLC) can be used to compose payment channels, therefore 
allowing the emergence of a second layer 
network for instantaneous and more anonymous payments: The Lightning Network. 

\medskip

The Lightning Network, LN from now on, was described in  \cite{PD15}. 
Its ambition is to scale to an instantaneous decentralized worlwide payment network with minimal fees, and 
is currently being tested and implemented.


In the original white paper \cite{PD15} the routing problem is only addressed  
in section 8.4 where only some tips are given about possible routing algorithms.
The belief is that routing tables are necessary for large operators as for current Internet routing 
algorithms as BGP or Cjdns protocols.
A proposal for implementation was presented in \cite{POOSZ16}. We refer to this article for historical 
background on the routing problem and ideas proposed and discussed among developers. In \cite{POOSZ16}
an algorithm called ``Flare'' is proposed that uses routing tables and beacon nodes that have a 
richer information about the geometry of the LN network.


The existence of a group of beacon nodes (even when selected randomly and renewed regularly) presents a threat 
to decentralisation. Indeed, even the global knowledge of the geometry of the network can be a vector of attack.


The main challenge is to implement a resilient, anti-fragile, secure, anonymous, scalable 
and decentralized routing algorithm. 


A necessary condition for perfect decentralisation is to have all nodes performing the same tasks and having access 
to the same information. Therefore we want to avoid ``beacon nodes'' as well as any privileged role of a node.


The main inspiration for our algorithm comes from ant behavior. 
Ant colonies exhibit a superior performance in their ``food finding'' algorithms. 
Mathematically speaking, they employ a ``balayage technique'' of the geometry of their natural habitat
in order to collect food. Only the best performers of these algorithms have survived 
natural selection over millions of years. This is an important reason 
to consider seriously these algorithms from a mathematical and geometrical point of view. 
Important research has been conducted by biologists, but also by other scientists and  
mathematicians. For general background we refer the reader to \cite{DS04}\footnote{Even in section 6 we can find 
a discussion on ``ant routing algorithms'' that, although more complicate, have the same flavor and share some  
ideas with the algorithm  presented in this article.}. 
One of the distinctive idea of ant algorithms is the 
use of pheromones in order to leave a ``trace'' of their passage and marking the food collecting routes. 
A natural routing algorithm is at work. The idea of ``leaving a trace'' appears 
as well in greek mythology in the myth of Ariadne's thread in order to solve in a practical efficient way
the ``path exiting problem'' in a labyrinth. 


We present in this article a routing algorithm where we implement similar ideas in order to achieve 
maximal decentralization. We present here only a first draft of the algorithm.
Further numerical 
simulations are necessary to evaluate the performance and scalability of the algorithm.


An interesting feature of the algorithm is that it has potential learning capabilities: It can adjust to the 
dynamic geometry of the network and can improve its performance over time (see \ref{sec_selfimprove}).

\section{Preliminaries on the Lightning Network.}

As described in the white paper \cite{PD15} the Ligthning Network (LN) is an aggregation of payment 
channels that can be composed. For the purpose of the present article we don't care about the specifics of 
the payment channels (if they are bidirectional or only one-way channels, etc). The payment channel network 
is assumed to be richly and randomly connected, as for the Bitcoin network.

\medskip

We assume that on top of the LN there is a separate, but richer, fast communication network. 
This means that a pair of participants (nodes of the network)
with an open payment channel also do have a communication channel, but pair's of participants 
without a direct payment channel can have a communication channel. This will be the case 
between Alice and Bob when both are part of the LN and want to make a transaction from Alice to Bob.
The route finding algorithm operates at the level of the communication network. The speed of communications
is one of the main bottlenecks for the speed of the LN.

\medskip

What we need to know and that we will use from the LN is that payment channels can be composed, once we know a 
payment path. Also payment channels support a maximal volume per transaction. This is a particular feature 
of the LN and it is discussed in Section \ref{sec_volume}.
We postpone a full analysis of possible misbehaviors and attacks on the LN, that deserves by itself a lengthly 
and careful discussion.

\section{Simple ant routing algorithm.}\label{sec_algo}

Alice wants to pay Bob a certain amount in Bitcoin through the LN network by composing already 
existent payment channels. We can assume that the network is 
composed of bidirectional channels (although this is not necessary as discussed in Section \ref{sec_hybrid})
and we can even have a mixture of unidirectional or bidirectional channels). 
We assume that a communication channels are open between neighboring nodes. 
In this section we describe 
how the network finds a path between Alice and Bob. This is only, for now, the simplest 
geometric problem.
In this section we don't address the limitation 
by the volume of the channels (you can assume that all channel are of 
a volume larger than the payment), nor we address the fee question (assume free transmission). 
Later on we consider channel payment volume limitations and fees (see Sections \ref{sec_fees} and \ref{sec_volume}). 

\begin{enumerate}
 \item Alice and Bob agree on a large random number. For example, Alice and Bob choose a 
 random $128$ bit numbers, $R(A)$ and $R(B)$ and exchange them in a secured way.
 \item Alice concatenates the bit $0$, and the hash\footnote{For example $h$ can be the commonly 
 used $\text{SHA256}$ in Bitcoin protocol. We take a hash in order to preserve a shared secret by Alice and Bob
 that they can used to prove that they are the originators of the transaction and can 
 serve extensions of the algorithm.} $R=h(R(A)^\frown R(B))$ to get a pheromone seed $S(A)=0^\frown R$ 
 and communicates $S(A)$ to its immediate neighbors in the LN with whom
 she has an open payment channel.
 \item Bob concatenates the bit $1$, $R(A)$ and $R(B)$ to get a pheromone seed 
 $S(B)=1^\frown R$ and communicates it to its neighbors in the LN with whom
 he has an open payment channel.
 \item Alice waits from an answer from its neighbors indicating her that a path has been found by the network.
 \item Bob waits to have news from Alice that a path has been found.
\end{enumerate}

If $S$ is a pheromone seed, we denote $S'$ the ``derived seed'' without the appended first bit, that is, 
the hash $R$. 
(thus $S= 0^\frown S'$ or  $S= 1^\frown S'$). If 
$S=0^\frown S'$ (resp. $S=1^\frown S'$) we denote by $\bar S$ 
the ``conjugate'' seed $\bar S =  1^\frown S'$ (resp. $\bar S =  0^\frown S'$).

\medskip

The nodes perform the following tasks (on top of a possible payment task if they are Alice or Bob).

\begin{enumerate}
 \item Each nodes reserves a fast access memory space for the routing tasks. 
 We refer to this as the ``LNmempool'' of the node\footnote{Not to be confused with Bitcoins's network's nodes mempool.}.
 \item Each node keeps in memory a numbered list of neighbors in the LN together with the relevant 
 information about its payment channel(s) opened with them. Also about historical performance of payments 
 through these neighbors.
 \item When a node receives a pheromone seed $S$, it checks if $S'$ is not a derived seed of a 
 seed already stored in the LNmempool.
 \item If $S'$ is not found, then it stores $S$ in the LNmempool together with the information 
 about the neighbor that has communicated $S$ (the ``transmitter neighbor''). 
 Then it broadcasts $S$ to the other neighbors.
 \item  If $S'$ is found, then it checks if $S$ is stored. 
 \begin{enumerate}
 \item If $S$ is stored it adds the information about the new transmitter neighbor.
 \item If $S$ is not stored it means that $\bar S$ is stored, so a matching occurs.
 \end{enumerate}
 \item  When a matching occurs, the node concatenates the bit $0$ to $S$ (resp. $\bar S$) and 
 constructs a ``matched seed'' $S_m= 0^\frown S$ (resp. $\bar S_m= 0^\frown \bar S$ 
 and sends it to the neighbors from which it received $S$ (resp. $\bar S$). 
 Note that ``matched seeds'' are one bit longer 
 than pheromone seed. The node keeps track of the neighbors having transmitted the unmatched seed. 
 \item When a node receives a matched seed $S_m$ it broadcasts it back to the 
 neighbors that send to him the unmatched seeds and keeps track of them. 
\end{enumerate}

Following this procedure, Alice will receive back several matched seeds that will correspond 
to different possible payment paths. She chooses one, say $S_m$, and concatenates 
a further $0$ bit creating a confirmed seed $S_c=0^\frown S_m$ that she sends back to 
the neighbor that send her the matched seed. Confirmed seeds are one bit longer than matched seeds, and two bits
longer than pheromone seeds. The nodes broadcast back the confirmed seed to the 
corresponding neighbor from which he received the matched seed until they reach the node that 
did the match. This node continues the broadcasting back of the confirmed seed until it reaches Bob. 
Once Alice receives from Bob the confirmation of the payment path, she starts the conditional payment chain as described 
in \cite{PD15} and the transaction will be completed.

Once the payment have been done, the nodes erase the data corresponding to the confirmed seed. 
Also, after some threshold time $\tau$, the nodes drop all data (matched and unmatched) older than $\tau$.
If no path is found after this threshold time, the ``path finding'' request of Alice and Bob is erased 
from the network. In that way the LNmempool keeps a controlled size. 
Each node can decide its own threshold time.

\section{Precisions on the simple routing algorithm.}

The simplified version described in the previous section needs to be made more precise for a proper transmission back of 
matched and confirmed seeds. It is important that the route gets ``locked'' so that an intermediate node does not mix different 
confirmed and matched seeds corresponding to different matchings. For this reason we introduce an Id of the matching. 
Also we want to minimize the information broadcast to the whole network. With this in mind,
we propose:

\begin{enumerate}
 \item We append a counter to pheromone seeds. Each time the pheromone seed is transmitted the counter is increased by $1$.
 \item Nodes only rebroadcast pheromone seeds if the counter is lower than previously seen for this pheromone seed.
 \item When a matching occurs, the node matching the pheromone seeds appends a ``matching Id'' which is a random number.
 \item Nodes broadcast back the matched seed so that the counter decreases and store the matching Id and the corresponding neighbors in the path. 
 \item Alice will confirm matched seeds and the path will be confirmed using the matching Id.
\end{enumerate}

Thus when a confirmed seed is received, the Id allows to choose the only next node in the path without ambiguity. 
Also, the counter restriction makes that the amount of pheromone seed data broadcast to the network is reduced. 
The counter provides some information about the length of the path which is not suitable. This happens if
the counters is started at $0$, but we may decide to start it at some random number to conceal that information,. Also we can add a random positive number in some range and 
larger than $1$ at each step to conceal even more this information. 

A node may be tempted to cheat on the counter making it smaller in order to sabotage or to increase 
its chances to be selected in the payment path. This may end-up creating 
a loop not reaching Alice or Bob, and, if repeated, the neighboring nodes may mark this node as ``non efficient'' (see Section \ref{sec_selfimprove} on 
how to select the most efficient neighbors using the historical data). Also Alice could require the fee to be proportional 
to the counter (more on fees in Section \ref{sec_fees}), which will force this sabotage strategy to cost on fees to the cheating node. 
There are also simple procedures that can be implemented by doing a final round 
in order to check that nodes are not cheating on the counter. For example, when Alice confirms 
the seed and sends it back to Bob through the path, each
node will append a random number, and the full sequence of random numbers reaches Bob. Bob checks that the number of random number 
matches the counter and communicates the whole sequence directly to Alice. 
Then Alice sends the whole sequence to the first node in the path that checks that his random number is the first one, deletes it and sends the remaining sequence 
to the second node that checks that his random number is the first remaining one, etc. 
The only thing that the cheater node can do is to remove some of the random numbers in order to dissimulate his cheat 
on the counter to Bob. 
But then in the final round this will be spotted by the first node whose random number was deleted\footnote{An exception occurs when the cheating node fails to 
increase the counter and does not add a random number in the final check. But then the cheating node acts ``transparently''.}

\section{Properties of the algorithm.}\label{sec_properties}

Some of the properties of the algorithm are listed below:

\begin{enumerate}
 \item Anonimity: Intermediary nodes have no information about Alice and Bob. 
 They only have information about the two neighbors in the payment path.
 \item Anonimity: No records of payments are kept.
 \item Anonimity: No node out of the payment path knows about the payment.
 \item Anonimity: No global information about the geometry of the network nor routing tables needed. 
 Computation load related to updates on routing tables is avoided. Nobody 
 needs to share information about their neighbors.
 \item Decentralization: All nodes perform the same function and follow the same rules. 
 In particular, no beacon nodes, and again no routing tables. 
 \item Scalability: Pending of numerical simulations, the routing algorithm should be able 
 to handle thousands of transactions per second.
 \item Instantaneous: Depending on the speed of communication.
\end{enumerate}

As drawback we can cite the intensive computational communication and processing which is typical from 
a totally decentralized protocol. As described in Section \ref{sec_selfimprove}, nodes can improve the efficiency and 
limit the workload by algorithmically selecting preferred neighbors from historical data.

\section{Channel volume compatibility.}\label{sec_volume}

The amount of the transaction should be compatible with the maximal volume of each payment channel (in the 
intended payment direction). Thus, some of the paths found could be incompatible with the payment amount.

\medskip

In order to ensure that the path found is volume compatible, Alice and Bob add an ``amount field'' to their 
pheromone seed. Then they broadcast it as before. Nodes in the network only broadcast the pheromone seed to 
neighbors which whom they have a volume compatible open payment channel. The rest of the procedure is the 
same and we get the subset of paths that are ``volume compatible'' with the transaction.

\medskip

Disclosure of the amounts is not suitable for the sake of preservation anonymity of the transaction. One may 
think to obfuscation implementations (as for the mimblewimble protocol), but for the LN this is not as important as for 
standard bitcoin transactions since Alice and Bob may issue multiple transactions dividing the total amount
into micropayments.

\section{Fee considerations.}\label{sec_fees}

The incentive for nodes to participate in payment paths in the LN are fees for transaction relay. Each node
freely determines the fee it will take to participate in a payment path. It will deduct the fee amount from the 
transacted amount. 
There are different approaches for the fee treatment. We describe one that  is simpler and straightforward, 
but leaks some 
information about the length of the payment path (that is correlated to the fee). One can
obfuscate the fee amount to intermediary nodes, but it involves a longer procedure.

\medskip

\textbf{Fee algorithm.}
In Alice and Bob's pheromone seed a ``maximum fee field'' and a ``current fee field'' are added. 
Alice and Bob initialize the ``current fee field'' by $0$ and the ``maximum fee field'' by  
the maximum amount that Alice is disposed to pay (that she communicates to Bob). 
Each time a pheromone seed is broadcast by a network node, it checks 
that the current amount and its fee is smaller than the maximal fee amount, it adds its fee to the amount 
in the current fee field. It keeps track in the LNmempool of the neighbor nodes it has communicated the pheromone
seed as well as the corresponding fee amounts. 

\medskip

When the matching node receives Alice and Bob pheromone seeds, it checks that the sum of the amounts in both 
``current fee fields'' is smaller than the maximum amount fee deducted from its own fee. Only under that 
provision the matching occurs. The total fee becomes then the sum of the current fee amounts 
of both pheromone seeds plus 
the matching node fee. Then in the matched seed this total fee amount is included in the ``current fee field''. 
When nodes broadcast the matched seed they do not change the ``current fee field``. The 
subsequent nodes that relay the matched seed relay respecting the fee they did indicate prior. If some node 
intents to increase the fee, it will be noticed by other nodes, an anomaly will be detected, 
and the matched seed will be not be further relayed.

\medskip

Alice will then received a list of proposed matched seeds with associated fee amounts. She can then select the lowest 
fee path (if she wishes).

\medskip

Obviously, in this setup the matching node can take advantage of his position and 
set a maximal fee in order to match the maximal fee. This could be a reward to be a matching node, but it is 
not clear that it has any advantage in trying to maximize its profit since after all there is a competition
between the paths discovered and Alice will probably select the paths with lowest fee, thus maximizing the fee
will result in lower probability for being selected.

\section{Hybrid  channels.}\label{sec_hybrid}

So far, the only property we have used of payment channels is its transitivity, i.e. that they can be composed. 
The algorithm described works as well for unidirectional channels, and one may even imagine an hybrid LN with 
swaps between different blockchains, as long as they can be composed. The algorithm is 
independent of the nature of the payment channels as long as they have the usual properties 
for composition of payment channels.

\section{Self-improvement of the algorithm.}\label{sec_selfimprove}

One typical feature of ant path finding algorithms is the self-reinforcement of paths by the intensity of the 
pheromone trace. This intensity increases with the number of ants taking the path. 
We can propose a similar reinforcement mechanism to boost the performance of the network. 

\medskip

This can be done with each node assigning a performance benchmark to each neighbor. In the neighbor tables with 
information about its payment channels, it can store historical information, as for example how many 
payments have been completed, or what is the total historical volume having circulated through that channel, etc
Short term and long term ''neighbor performance`` is an important data. 

\medskip

Then each node can selectively broadcast the pheromone seeds according to the numerical criteria he wishes
to implement taking into account its neighbor's numerical data. 
In particular, the comparison between short term and long term data reflects the dynamical changes of the 
network and allow the nodes to adjust to the new geometry.
The nodes may also use random algorithms to select their preferred neighbors for broadcasting pheromone seeds. 
A Pareto type distribution in terms of an historical activity index will emulate closely the ant algorithms (see
\cite{FVV13} for example).

\medskip

The historical data can be used not only to detect the best performers neighboring nodes, but  also neighboring 
nodes that misbehave and sabotage the LN operations. In particular the ratio between pheromone seeds, and matched seeds 
transmitted and final payment paths realized is an important data to detect local misbehavior.


\newpage
\begin{thebibliography}{1}


\bibitem[1]{DOR18} C.~Decker, R.~Russell, O.~Osuntokun. \textit{eltoo:A simple layer2 protocol for Bitcoin}, 2018.


\bibitem[2]{DW15} C.~Decker, R.~Wattenhofer. \textit{A fast and scalable payment
network with bitcoin duplex micropayment channels}, Symposium on Self-Stabilizing Systems, 2015.

\bibitem[3]{DS04} M.~Dorigo, T.~St\"utzle. \textit{Ant colony optimization}, 
MIT Prees, 2004.


\bibitem[4]{FVV13} M.~Vela-P\'rez, M.A.~Fontelos,J.J.L.~Velázquez. \textit{Ant foraging and 
geodesic paths in labyrinths: Analytical and computational results}, Journal of Theoretical 
Biology 320, p.100–112, 2013


\bibitem[5]{HS15} M.~Hearn, J.~Spilman. \textit{Bitcoin contracts}, \texttt{en.
bitcoin.it/wiki/Contracts}, 2015.


\bibitem[6]{N08} S.~Nakamoto. \textit{Bitcoin: a peer-to-peer electronic cash system}, 
\texttt{Bitcoin.org}, 2008.

\bibitem[7]{PD15} J.~Poon, T.~Dryja. \textit{Lightning network}, 2015.

\bibitem[8]{POOSZ16} P.~Prihodko, A.~Ostrovskiy, O.~Osuntokun, M.~Sahno, S.~Zhigulin. 
\textit{Flare: An Approach to Routing in Lightning Network}, version 1.0, 2016.

  

\end{thebibliography}
\end{document}